\documentclass[twocolumn,prd,nofootinbib,eqsecnum]{revtex4}
\usepackage{graphicx}
\usepackage[utf8]{inputenc}
\usepackage{textcomp}
\usepackage{amssymb}
\usepackage{amsmath}

\newcommand{\beq}{\begin{equation}}
\newcommand{\eeq}{\end{equation}}
\newcommand{\beam}{b} 

\usepackage{graphicx}
\usepackage{amsmath}

\newcommand{\tr}{{\rm Tr}}
\newcommand{\Gammabold}{{\boldsymbol\Gamma}}

\begin{document}
\title{Fisher Matrix Optimization of Cosmic Microwave Background Interferometers}
\author{Haonan Liu}
\email{haonan.liu@richmond.edu}
\author{Emory F. Bunn}
\affiliation{Physics Department, University of Richmond, Richmond, VA  23173}

\begin{abstract}
We describe a method for forecasting errors in interferometric
measurements of polarization of the cosmic microwave background (CMB)
radiation, based on the use of the Fisher matrix calculated from the
visibility covariance and relation matrices.  In addition to noise and
sample variance, the method can account for many kinds of systematic
error by calculating an augmented Fisher matrix, including parameters
that characterize the instrument along with the cosmological
parameters to be estimated. The method is illustrated with examples of
gain errors and errors in polarizer orientation. The augmented Fisher
matrix approach is applicable to a much wider range of problems beyond
CMB interferometry.
\end{abstract}

\maketitle

\section{Introduction}

Observations of the cosmic microwave background (CMB) are among the chief drivers of the revolutionary advances in cosmology over the past 20 years (see, \textit{e.g.}, \cite{hureview,challinorreview} and references
therein), and further observations, particularly of CMB polarization, are likely to be of major importance in the coming years. CMB polarization can probe the early Universe in a wide variety of ways, including possibly providing direct evidence of a stochastic gravitational wave background, which would give direct evidence in support of inflation \cite{zalsel,selzal,kkslett,kks}.

The key to analyzing CMB polarization data is the separation of the signal into a scalar $E$ component and a pseudoscalar $B$ component. The inflationary signal is sought in the $B$ component. Because this component is predicted to be significantly weaker than the $E$ component (and both are much weaker than the temperature anisotropy), detection of $B$ modes is a daunting task.

Control of systematic errors will be vital to the success of the quest to characterize B modes. In particular, some errors may cause ``leakage'' into the $B$ signal from the much larger $E$ and temperature anisotropy ($T$) signals. (Such leakage occurs even in the absence of systematic errors. See, \textit{e.g.}, \cite{LCT,bunn,bunnerratum,bunnetal,bunneb,kimeb,bowyereb,lewiseb,parkng,smitheb1,smitheb2,smithzaleb,caofang,zhaoeb}.)
In considering the design of future instruments, methods of assessing the severity of various sources of error are quite valuable. Detailed simulations ({\it e.g.}, \cite{karakcibayes,zhangml,karakcisys,sutterbayes2012}) often provide the best assessment, but approximate analytic methods ({\it e.g.}, \cite{HHZ,bunnsys}) can provide valuable insight before undertaking the computational effort of a full simulation.

In this paper, we describe a Fisher-matrix method for assessing the ability of interferometric instruments to detect CMB polarization signals, with particular emphasis on the effect of systematic errors. Interferometers have been important in CMB science in the past, including making some of the early detections of CMB polarization \cite{readhead,leitch}. 
The formalism for analyzing interferometric CMB observations has been well-developed \cite{HobLasJon,HobMag,WCDH,HobMais,Mye,BW}
Although most current and planned instruments are traditional imaging telescopes, at least one interferometer is under development \cite{qubicpiat,qubicghribi}. Interferometers and imaging telescopes have quite different susceptibility to various sources of error, so it would seem worthwhile to develop tools to understand both approaches.

The method described in this paper uses Fisher matrices to forecast the errors on cosmological parameters obtainable from a given instrument. After developing the general method for calculating Fisher matrices and the resulting error forecasts for an ideal (systematic-error-free) interferometer, we extend the method to allow consideration of systematic effects. In particular, we calculate an ``augmented Fisher matrix,'' in which we treat parameters describing the various systematic effects as unknowns to be estimated along with the cosmological parameters of interest. The resulting Fisher matrices allow us to forecast the errors expected on cosmological parameters, taking into account our ignorance of the systematic error parameters.

Although we apply our formalism to the specific case of CMB polarization measurements, the approach we describe has broader applicability. In particular, upcoming efforts to map the Universe through 21-cm tomography
(see, \textit{e.g.}, \cite{pritchard21cm,morales21cm} and references therein) involve the extraction of faint signals from interferometric data sets and hence have significant overlap with CMB polarization. It would be of interest to explore the generalization of our approach to this class of observations.

The remainder of this paper is organized as follows. Section \ref{sec:formalism} lays out the formalism for calculating covariance matrices and Fisher matrices for interferometers, including systematic effects. Section \ref{sec:analytic} reviews the approximate analytic treatment of systematic effects in ref.~\cite{bunnsys}. Section \ref{sec:results} presents the results of a variety of tests of the method, and Section \ref{sec:discussion} provides a brief discussion.

\begin{figure*}[t]
\includegraphics[width=2.8in]{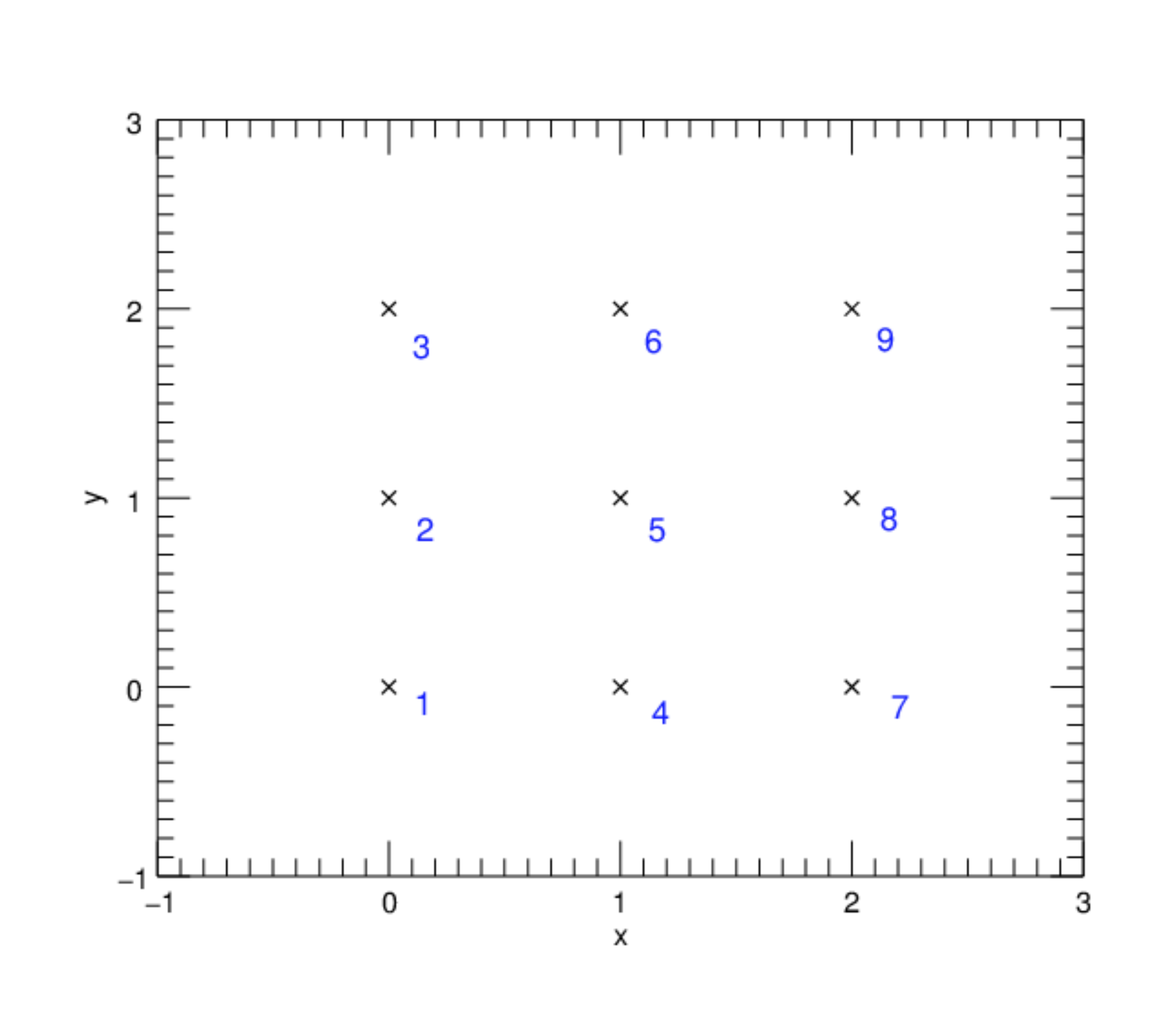}
\includegraphics[width=4.2in]{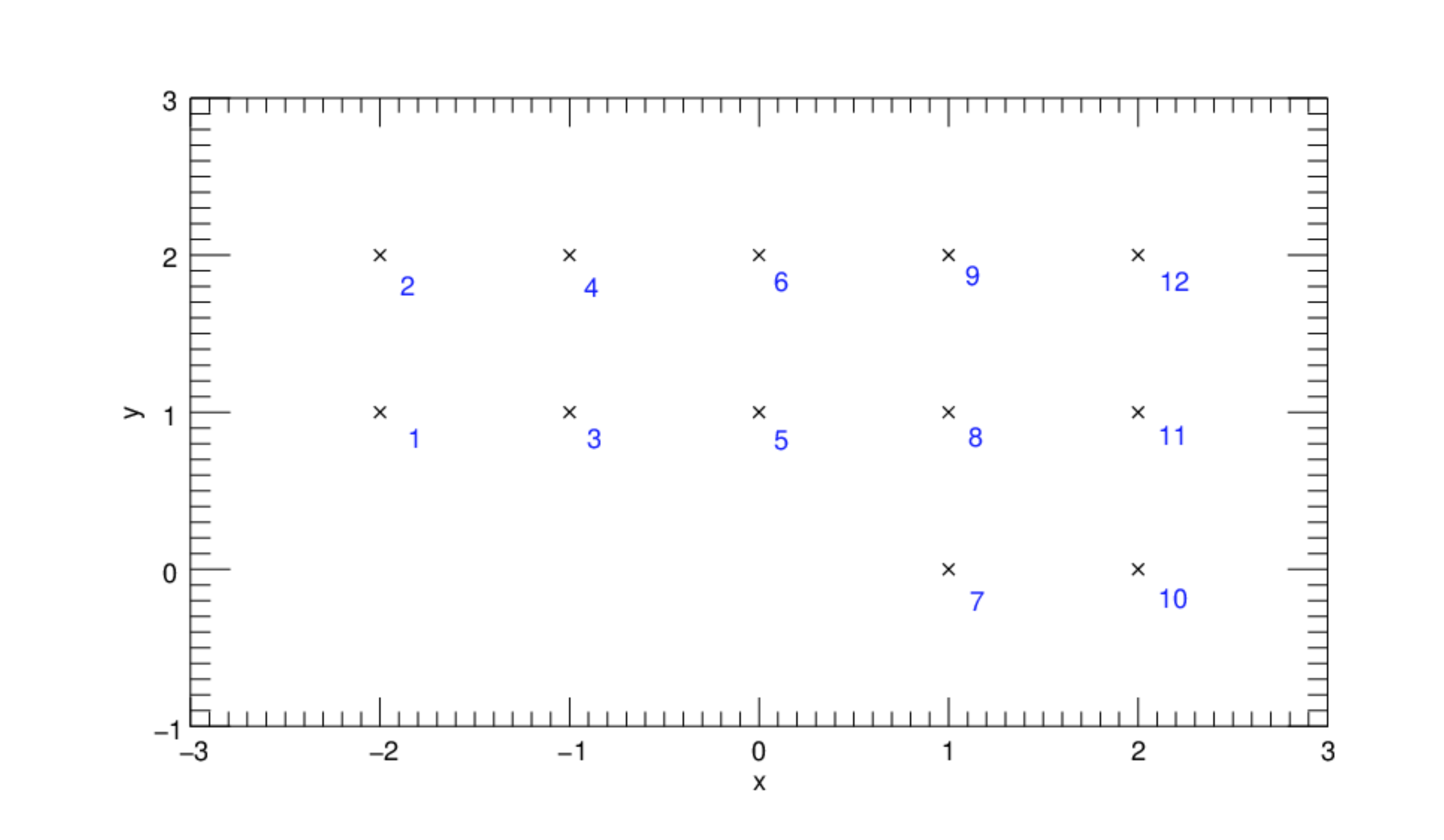}
\caption{A close-packed square array and its corresponding baseline graph with $N_a=3$.}
\label{fig:squarearray}
\end{figure*}

\section{Formalism}
\label{sec:formalism}
\subsection{Visibilities}
\label{subsec:vis}
Consider a close-packed square array with $N_a$ antennas on each side. For antennas $j$ and $k$ that are separated by the vector 
$\delta\mathbf{r}=(x,y)$ on the antenna map, the baseline $\mathbf{u}$ has the form 
\begin{equation}
\mathbf{u}=\delta \mathbf{r} \Delta u=(x\Delta u,y\Delta u).
\end{equation}
Here $x$ and $y$ are any integers that satisfy $-(N_a-1)\leqslant x,y\leqslant N_a-1$. 
The baseline for two adjacent antennas is $\Delta u=D/\lambda$, where $D$ is the antenna diameter. We will assume diffraction-limited antennas, for which the Gaussian beam width is $\beam=0.518\lambda/D$, so that
$\Delta u=0.518/\beam$.  

The total number of independent baselines is $\frac{1}{2}[(2N_a-1)^2-1]$, 
where the $-1$ appears
because we do not include the case $x=y=0$, and the $\frac{1}{2}$ is
included to avoid redundancy since
$V(\mathbf{u})=V^*(-\mathbf{u})$. We include all baselines with $y>0$
and all with $y=0$ and $x\ge 0$, labeling them in the order shown in Figure
\ref{fig:squarearray}. 

Now, suppose a monochromatic plane wave with angular frequency $\omega =2\pi c/\lambda$ that approaches the center of one antenna from the direction $\hat{\mathbf{r}}$ has the form
\begin{equation}
\mathbf{E}=\boldsymbol{\epsilon}(\hat{\mathbf{r}})e^{i(\omega t-\mathbf{k}\cdot\mathbf{x})},
\end{equation}
where $\mathbf{k}$ is the wavevector defined by $\mathbf{k}=-(2\pi/\lambda)\hat{r}$. Then the $2\times 2$ matrix of visibilities corresponding to a given baseline $\mathbf{u}$ is \cite{bunnsys}
\begin{equation}
\mathbf{V}(\mathbf{u})=\int{\mathbf{S}(\mathbf{r})A(\mathbf{r})e^{2\pi i \mathbf{u}\cdot \mathbf{r}} d^2 \mathbf{r}},
\end{equation}
where the Stokes matrix $\mathbf{S}$ is given by
\begin{equation} 
\begin{split}
{\bf S}&=
\begin{pmatrix}
S_{XX} & S_{XY}\\
S_{YX} & S_{YY}
\end{pmatrix}\\
&=
\begin{pmatrix}
I+Q & U+iV\\
U-iV & I-Q
\end{pmatrix},
\end{split}
\label{eq:stokes}
\end{equation}
and $A(\mathbf{r})$ is the antenna pattern, which we will take to be
Gaussian,
\begin{equation}\label{antennaPattern} 
A(\mathbf{r}) =  e^{-\mathbf{r}^2/{2\beam^2}}
\end{equation}
with $\beam$ the beam size.
Note that we are considering experiments that combine linear polarization states $(X,Y)$, not circular polarization states.

We define Stokes $Q$ and $U$ visibilities,
\begin{eqnarray} \label{vqvu}
V_Q\equiv V_Q(\mathbf{u})=\int{Q(\mathbf{r})A(\mathbf{r})e^{2\pi i \mathbf{u}\cdot \mathbf{r}} d^2 \mathbf{r}},\\ 
V_U\equiv V_U(\mathbf{u})=\int{U(\mathbf{r})A(\mathbf{r})e^{2\pi i \mathbf{u}\cdot \mathbf{r}} d^2 \mathbf{r}}.
\end{eqnarray}
As usual in CMB studies, we will
express Stokes parameters and their associated visibilities
in thermodynamic temperature units.

By the convolution theorem,
\begin{eqnarray}
V_Q=(2\pi)^2\int{\int{\tilde{Q}(\mathbf{k})\tilde{A}^*(\mathbf{k}+2\pi \mathbf{u})  d^2 \mathbf{k}}},\label{eq:vqfourier}\\
V_U=(2\pi)^2\int{\int{\tilde{U}(\mathbf{k})\tilde{A}^*(\mathbf{k}+2\pi \mathbf{u})  d^2 \mathbf{k}}},\label{eq:vufourier}
\end{eqnarray}
where the tilde denotes a two-dimensional Fourier transform in the 
flat-sky approximation. 

From equation (\ref{eq:stokes}), we can see that the $Q$ and $U$ visibilities are related to the visibility matrix $\mathbf{V}$ by
\begin{eqnarray}
V_Q=\frac{1}{2}(V_{XX}-V_{YY}),\label{eq:vq}\\
V_U=\frac{1}{2}(V_{XY}+V_{YX}).\label{eq:vu}
\end{eqnarray}
In practice, we do not usually use (\ref{eq:vq}) to measure $V_Q$, as it involves cancellation of the large contribution from Stokes $I$. Instead,
we usually measure $V_Q$ by rotating the axes for our polarization basis
through 
45\textdegree\ and then using equation (\ref{eq:vu}).

The Stokes parameters are related to the E and B modes in the following way:
\begin{align}
\tilde{Q}(\mathbf{k})&=\tilde{E}(\mathbf{k})\cos(2\phi)+\tilde{B}(\mathbf{k})\sin(2\phi),\label{q def}\\
\tilde{U}(\mathbf{k})&=-\tilde{E}(\mathbf{k})\sin(2\phi)+\tilde{B}(\mathbf{k})\cos(2\phi)\label{u def},
\end{align}
where $\phi$ is the angle between the vector $\mathbf{k}$ and the $x$ axis.
We assume that the $E$ and $B$ contributions are independent, statistically isotropic random fields with the flat-sky power spectra $P_E,P_B$, which are defined as $P_{E,B}(k)=C_l^{E,B}$ with $l=k$. Then  
\begin{equation}
\langle \tilde{E}({\bf k}_1)\tilde{E}^*({\bf k}_2)\rangle=\frac{P_E(k_1)}{(2\pi)^2}\delta({\bf k}_1-{\bf k}_2),
\end{equation}
with a similar expression for B. The extra $(2\pi)^2$ factor is caused by the approximation of a flat-sky power spectrum from a full sky power spectrum.

We then have
\begin{align}\label{qq}
\langle\tilde{Q}(\mathbf{k}_1)\tilde{Q}^*(\mathbf{k}_2)\rangle&=\delta(\mathbf{k}_1-\mathbf{k}_2)/(2\pi)^2\notag\\
                                                            &[P_E(\mathbf{k})\cos^2(2\phi)+P_B(\mathbf{k})\sin^2(2\phi)],
\end{align} 
\begin{align}
\langle\tilde{U}(\mathbf{k}_1)\tilde{U}^*(\mathbf{k}_2)\rangle&=\delta(\mathbf{k}_1-\mathbf{k}_2)/(2\pi)^2\notag\\
                        &[P_E(\mathbf{k})\sin^2(2\phi)+P_B(\mathbf{k})\cos^2(2\phi)],
\end{align}
\begin{align}
\langle\tilde{Q}(\mathbf{k}_1)\tilde{U}^*(\mathbf{k}_2)\rangle&=\delta(\mathbf{k}_1-\mathbf{k}_2)/(2\pi)^2\notag\\
&[-\frac{1}{2}P_E(\mathbf{k})\sin(4\phi)+\frac{1}{2}P_B(\mathbf{k})\sin^2(2\phi)].
\end{align}
We will use these results in the next section for the deduction of the visibility covariance matrix.

For regular arrays of the sort considered here, multiple pairs of antennas can have the same separation vector $\delta{\bf r}$ and hence correspond to the same baseline.
We refer to the visibility measured from a given antenna pair as a ``micro-visibility'' and say that multiple micro-visibilities correspond to the same ``macro-visibility.''
Since all micro-visibilities corresponding to a given macro-visibility should be equal (in the absence of noise), we average them together to get the measured value of the macro-visibility.

To be specific, let $V_J$ represent the $J$th macro-visibility. Let
$S_J$ be the set of all micro-visibilities corresponding to that
macro-visibility, with each micro-visibility labeled by a pair of antennas
$(ab)$. Also let $N_J$ be the number of
micro-visibilities corresponding to that macro-visibility (i.e., the
cardinality of $S_J$). Then
\begin{equation} \label{macrov}
V_{XJ}={1\over N_J}\sum_{(ab)\in S_J}V_{X}^{(ab)},
\end{equation}
where $X\in\{Q,U\}$, and $V_X^{(ab)}$ labels a micro-visibility.

For instance, consider a $3\times 3$ array with antennas labeled from
1 to 9 as in Figure \ref{fig:squarearray}.
There are twelve macro-visibilities.
Using the numbering scheme in the figure,
$J=7$ corresponds to the macro-visibility $(1,0)$ -- that is, to
antennas that are separated by one unit in the $x$ direction and zero
units in the $y$ direction. Then
\begin{equation}
S_7=\{(14),(25),(36),(47),(58),(69)\},
\end{equation}
and of course $N_7=6$.

\subsection{Fisher Information Matrix}
\subsubsection{Real Values}


Although we will be working with complex data (visibilities), we begin by reviewing the Fisher matrix formalism for real data.

Let $\vec r=(r_1,\ldots,r_n)^T$ be a random vector drawn from a
multivariate normal distribution with mean zero and covariance matrix
$\Gammabold$ with elements
\begin{equation} \label{cov}
\Gamma_{jk}=\langle r_ir_j\rangle.
\end{equation}

The probability density for $\vec r$ is
\begin{equation} \label{fr}
f(\vec r)={e^{-{1\over 2}\vec r^T\Gammabold^{-1}\vec r}\over (2\pi)^{n/2}\det^{1/2}(
\Gammabold)}.
\end{equation}

Now suppose that $\Gammabold$ is a function of some set of $m$ parameters $\theta_1,\theta_2,\ldots,\theta_m$. These may be cosmological parameters which we 
would like to estimate from the data, but as we will see they can
also be parameters characterizing the experiment itself (e.g., the
gain on an antenna or the orientation of a polarizer).

We define the likelihood function
\begin{equation}
L(\vec \theta)=f(\vec r|\vec\theta).
\end{equation}
For convenience, define the log-likelihood function
\begin{equation}
{\cal L}=-\ln L={1\over 2}\vec r^T\Gammabold^{-1}\vec r+{1\over 2}\ln\det(\Gammabold)
+{n\over 2}\ln(2\pi).
\end{equation}
Then the Fisher Information Matrix is defined to have elements
\begin{equation}
F_{jk}=\left<{\partial^2{\cal L}\over\partial\theta_j\partial\theta_k}\right>,
\end{equation}
which can be shown to be
\begin{equation} \label{fisher}
F_{jk}={1\over 2}\tr\left(\Gammabold^{-1}{d\Gammabold\over d\theta_j}
\Gammabold^{-1}{d\Gammabold\over d\theta_k}\right).
\end{equation}

The Fisher matrix tells us the expected errors on the parameters. In particular, $1/\sqrt{F_{ii}}$ is the expected error on $\theta_i$ assuming all the other parameters are known, and $\sqrt{(F^{-1})_{ii}}$ is the optimal error on $\theta_i$ assuming
all other parameters are unknown (\textit{e.g}, \cite{berger}). 


\subsubsection{Complex Values}
We now consider complex vectors. Let $\vec z=(z_1,\ldots, z_n)^T$ be a
vector of complex Gaussian random
numbers with zero mean (e.g., a set of visibilities). 
One way to proceed is to think of this
$n$-dimensional complex vector as a $2n$-dimensional real vector. To
be specific, let $\vec x={\rm Re}(\vec z)$ and $\vec y={\rm Im}(\vec
z)$, and define
\begin{equation}
\vec r=\begin{pmatrix}\vec x\\ \vec y\end{pmatrix}
=\begin{pmatrix}x_1\\ \vdots \\ x_n \\ y_1 \\ \vdots \\ y_n\end{pmatrix}.
\end{equation}

Then define a $2n\times 2n$ covariance matrix 
\begin{equation} \label{gammar}
\Gammabold^{(r)}=\begin{pmatrix}
\Gammabold_{xx}  & \Gammabold_{xy}\\
\Gammabold_{yx} & \Gammabold_{yy}
\end{pmatrix},
\end{equation}
where the four $n\times n$ sub-matrices are $\Gammabold_{xx}=\langle \vec x\vec x^T\rangle$,
etc. Then all of the results of the previous section apply.

Often, it is inconvenient to express complex vectors in terms of the real and imaginary parts as above. Instead, we define two new matrices: the complex covariance matrix $\Gammabold$ and the complex relation matrix $\mathbf{C}$
with elements
\begin{equation}
\Gamma_{jk}=\langle z_jz_k^*\rangle,
\end{equation}
\begin{equation}
C_{jk}=\langle z_jz_k\rangle.
\end{equation}

It is straightforward to check that
\begin{align} \label{gamma}
\Gammabold &= \Gammabold_{xx}+\Gammabold_{yy}+i(\Gammabold_{yx}-\Gammabold_{xy})\\
\mathbf{C}&=\Gammabold_{xx}-\Gammabold_{yy}+i(\Gammabold_{yx}+\Gammabold_{xy}).\label{relation}
\end{align}

These two methods of approaching the complex Gaussian vectors are closely related. To see this relationship, define another $2n$-dimensional vector
\begin{equation}
\vec c=\begin{pmatrix} \vec z\\ \vec z^*\end{pmatrix}.
\end{equation}
Then 
\begin{equation}
\vec c = \mathbf{B}\vec r,
\end{equation}
where $\mathbf{B}$ can be written in block form as
\begin{equation}
\mathbf{B}=\begin{pmatrix}\mathbf{I} & i\mathbf{I} \\ 
\mathbf{I} & -i\mathbf{I}\end{pmatrix},
\end{equation}
and $\mathbf{I}$ is the $n$-dimensional identity matrix. 
Then
\begin{equation}
\Gammabold^{(c)}\equiv \langle \vec c \vec c^\dag\rangle
=\begin{pmatrix}\langle \vec z\vec z^\dag\rangle
& \langle \vec z\vec z^T\rangle \\
\langle \vec z\vec z^T\rangle^* & \langle \vec z\vec z^\dag\rangle^*
\end{pmatrix}
=\begin{pmatrix}\Gammabold & \mathbf{C} \\ \mathbf{C}^* & \Gammabold^*\end{pmatrix}.
\end{equation}
But we can also write
\begin{equation}
\Gammabold^{(c)}=\langle \vec c\vec c^\dag\rangle
=\mathbf{B}\langle \vec r\vec r^T\rangle \mathbf{B}^\dag.
\end{equation}
Setting these two expressions equal gives us 
\begin{eqnarray} \label{cr}
\Gammabold^{(c)}&=&\mathbf{B}\Gammabold^{(r)}\mathbf{B}^\dag,\\
\Gammabold^{(r)}&=&\mathbf{B}^{-1}\Gammabold^{(c)}(\mathbf{B}^\dag)^{-1}=
\frac{1}{4}\mathbf{B}^\dag\Gammabold^{(c)}\mathbf{B},
\label{eq:cr2}
\end{eqnarray}
using the fact that $\mathbf{B}^{-1}={1\over 2}\mathbf{B}^\dag$.
Equations (\ref{cr}) and (\ref{eq:cr2}) provide a convenient way of converting back and forth between the real ($\vec x,\vec y$) and complex $(\vec z,\vec z^*)$ representations of our random vector.

Equation (\ref{fr}) gives the probability density for a $2n$-dimensional real vector:
\begin{equation} \label{fr2n}
f(\vec r)={e^{-{1\over 2}\vec r^T(\Gammabold^{(r)})^{-1}\vec r}\over
(2\pi)^n\det^{1/2}(\Gammabold^{(r)})}.
\end{equation}
We can write this in terms of the complex vector $\vec c$
using $\vec r=\mathbf{B}^{-1}\vec c$:
\begin{equation} \label{fc}
f(\vec c)={e^{-{1\over 2}\vec c^\dag (\Gammabold^{(c)})^{-1}\vec c}\over 
\pi^n\det^{1/2}(\Gammabold^{(c)})}.
\end{equation}
The factor $2$ in equation (\ref{fr2n}) is missing in equation (\ref{fc}) because $\det(\Gammabold^{(c)})=|\det(\mathbf{B})|^2\det(\Gammabold^{(r)})$,
and $\det(\mathbf{B})=(-2i)^n$. 

The Fisher matrix elements are given by equation (\ref{fisher}), using 
$\Gammabold^{(r)}$ as the covariance matrix. If, as is often the case, it is easier to work with $\Gammabold^{(c)}$, we simply use equation (\ref{eq:cr2}) to relate
the two matrices.
We will apply these results to the vector of complex visibilities
in the next section.

\begin{figure*}[t]
\includegraphics[width=3.5in]{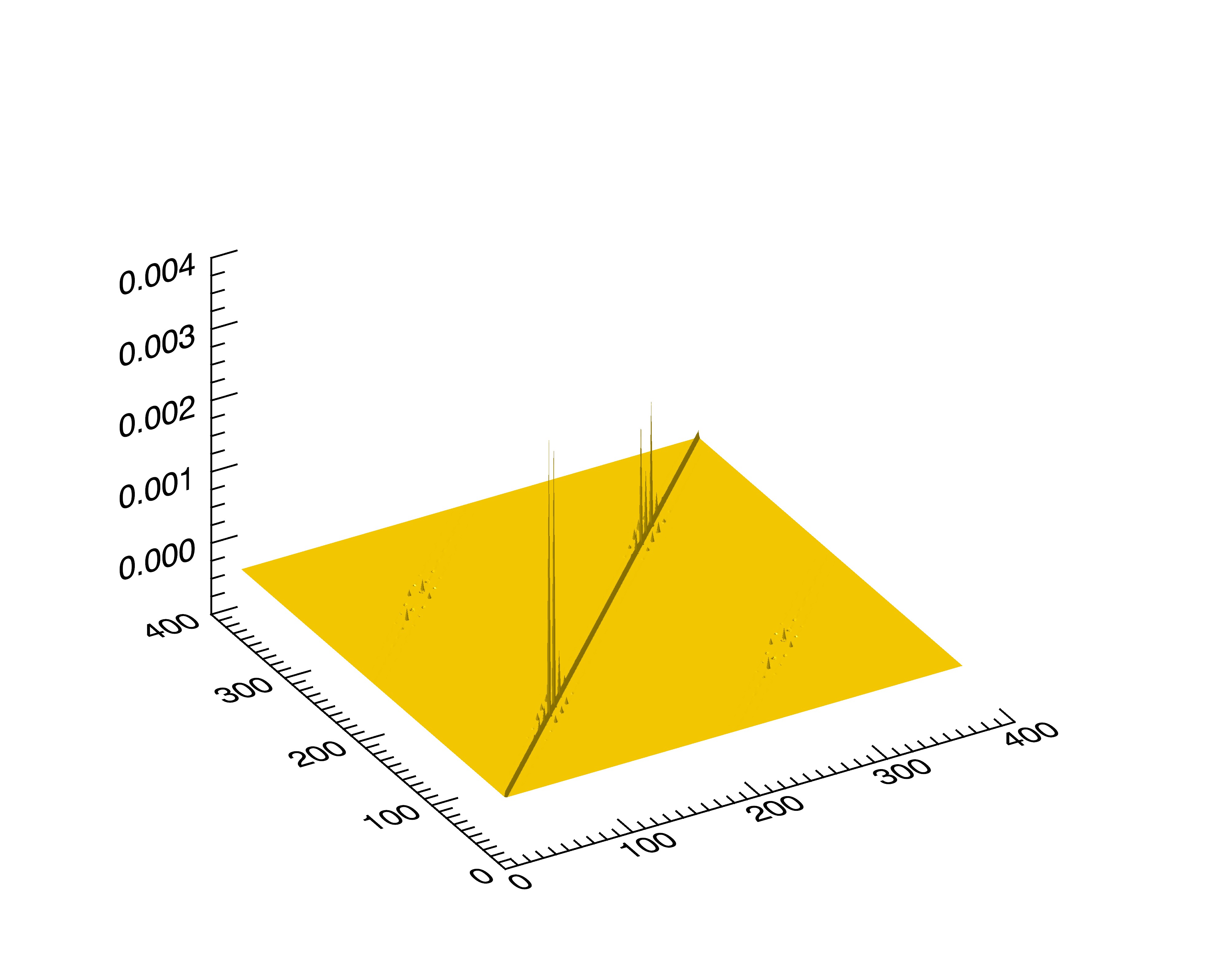}
\includegraphics[width=3.5in]{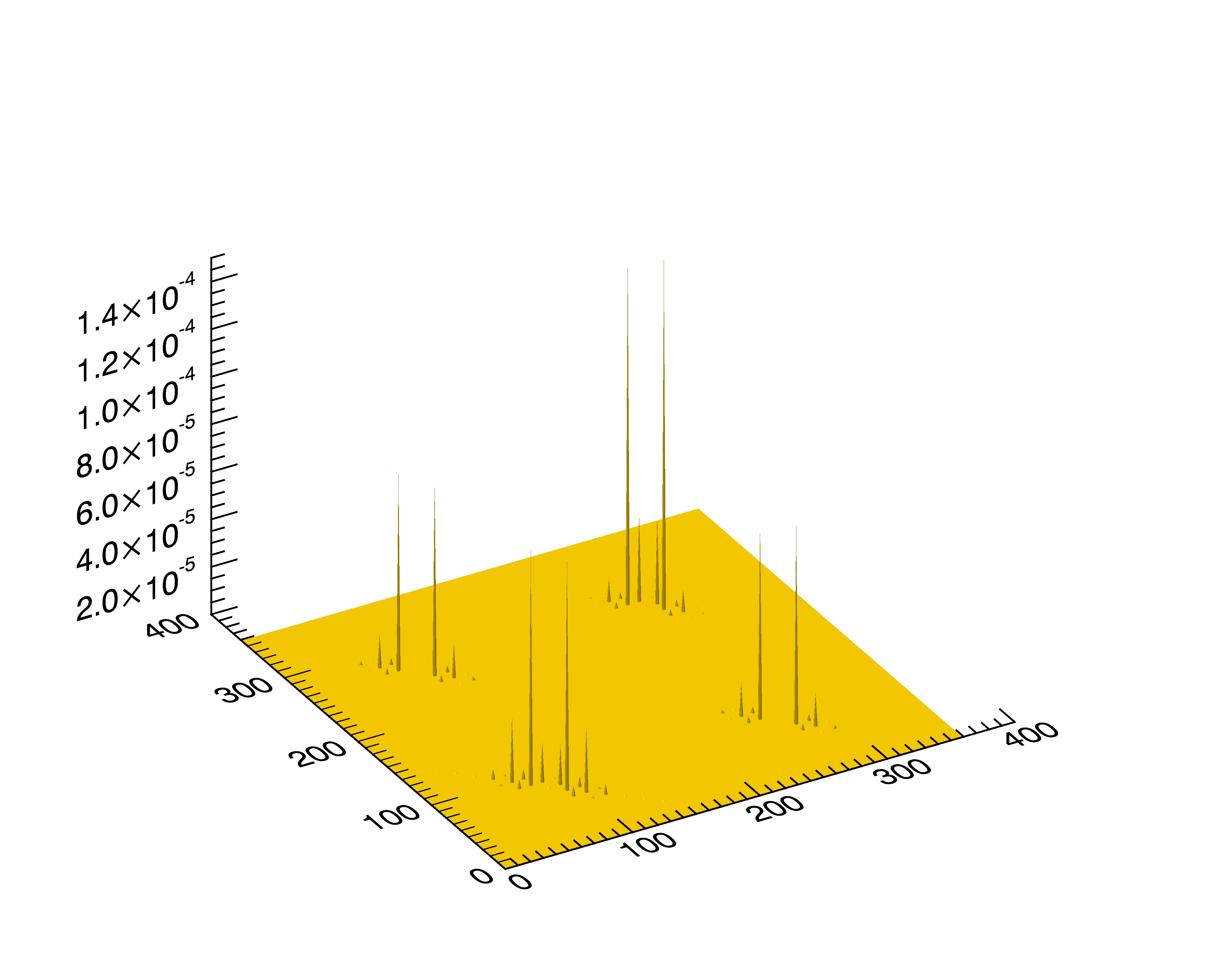}
\caption{Covariance and relation matrices $\Gammabold$ and $\mathbf{C}$
for an array with $N_a=10$, $\beam=0.0873$ rad, noise $n=0.01\,\mu$K. The sizes of both matrices are $360\times 360$ and the unit of the vertical axis is $\mu{\rm K}^2$. Notice that noise only affects the $\Gammabold$ matrix. }
\label{fig:matrices}
\end{figure*}

\subsection{Visibility Covariance Matrix and Fisher Matrix}

We now apply the results of the previous section to visibility data. Suppose that our data consists of a set of visibilities $\vec c = 
(V_Q(\mathbf{u}_1),V_Q(\mathbf{u_2}),\ldots,V_U(\mathbf{u}_1),V_U(\mathbf{u}_2),\ldots)$.
Consider an element of the upper left quadrant of the covariance matrix $\Gammabold$, which is the covariance of two $Q$ visibilities.
According to equations (\ref{eq:vqfourier}) and (\ref{qq}),
\begin{align}
& \qquad \langle V_Q(\mathbf{u}_1)V_Q^*(\mathbf{u}_2)\rangle \notag\\
&=(2\pi)^2\int{\langle\tilde{Q}(\mathbf{k}_1)\tilde{Q}^*(\mathbf{k}_2)\rangle}
\tilde{A}^*(\mathbf{k}_1+2\pi \mathbf{u}_1)\tilde{A}(\mathbf{k}_2+2\pi \mathbf{u}_2) d^2 \mathbf{k}\nonumber\notag\\
&=(2\pi)^2\int{(\delta(\mathbf{k}_1-\mathbf{k}_2)[P_E(\mathbf{k})\cos^2(2\phi)+P_B(\mathbf{k})\sin^2(2\phi)]}\notag\\
   &\qquad\tilde{A}^*(\mathbf{k}_1+2\pi \mathbf{u}_1)\tilde{A}(\mathbf{k}_2+2\pi \mathbf{u}_2) d^2 \mathbf{k}\notag\\
&=(2\pi)^2\int{[P_E(\mathbf{k})\cos^2(2\phi)+P_B(\mathbf{k})\sin^2(2\phi)]}\notag\\
   &\qquad\tilde{A}^*(\mathbf{k}+2\pi \mathbf{u}_1) \tilde{A}(\mathbf{k}+2\pi \mathbf{u}_2) 
d^2 \mathbf{k}.\label{vqvq}
\end{align}
As long as $A$ is isotropic, this expression and the corresponding ones for $\langle V_QV_U^*\rangle$ and $\langle V_UV_U^*\rangle$ can be expressed as a single integral by writing $d^2k$ in polar coordinates and integrating over the angular coordinate, as shown in the Appendix.

As long as the antenna pattern $A$ has reflection symmetry,
$\tilde{A}$ is real, and hence
$\left<V_Q(\mathbf{u}_1)V_Q^*(\mathbf{u}_2)\right>$ is real as
well. By similar reasoning,
$\left<V_Q(\mathbf{u}_1)V_U^*(\mathbf{u}_2)\right>$ and $\left<
V_U(\mathbf{u}_1)V_U^*(\mathbf{u}_2)\right>$ are also real. Hence we
conclude $\Gammabold$ is real for visibility data. The same is true for
the relation matrix $\mathbf{C}$. 

Since $\Gammabold_{xx}$, $\Gammabold_{yy}$, $\Gammabold_{yx}$ and $\Gammabold_{xy}$ are all real, equations (\ref{gamma}) and (\ref{relation}) imply that
$\Gammabold_{yx}-\Gammabold_{xy}=
\Gammabold_{yx}+\Gammabold_{xy}=0$, and hence that 
\begin{align}
\Gammabold_{yx}&=\Gammabold_{xy}=0,\\
\Gammabold_{xx}&=(\Gammabold+\mathbf{C})/2,\\
\Gammabold_{yy}&=(\Gammabold-\mathbf{C})/2.
\end{align}
Therefore, 
from the definition of $\Gammabold^{(r)}$ (\ref{gammar}), 
we see that $\Gammabold^{(r)}$ is a block matrix
\begin{equation}
\Gammabold^{(r)}=\begin{pmatrix}
\Gammabold_{xx}  & 0\\
0 & \Gammabold_{yy}
\end{pmatrix}
=\begin{pmatrix}
(\Gammabold+\mathbf{C})/2 & 0\\
0 & (\Gammabold-\mathbf{C})/2
\end{pmatrix}.
\end{equation}
An example of $\Gammabold$ and $\mathbf{C}$ can be found in Figure 
\ref{fig:matrices}.

Now we attain the final form for the visibility Fisher matrix
\begin{align}
F_{jk}={1\over 2}&\tr\left((\Gammabold_{xx})^{-1}{d\Gammabold_{xx}\over d\theta_j}(\Gammabold_{xx})^{-1}{d\Gammabold_{xx}\over d\theta_k}\right)\notag\\
&+{1\over 2}\tr\left((\Gammabold_{yy})^{-1}{d\Gammabold_{yy}\over d\theta_j}(\Gammabold_{yy})^{-1}{d\Gammabold_{yy}\over d\theta_k}\right)
\label{eq:finalfisher}
\end{align}

\subsection{The Augmented Fisher Matrix}
\label{sec:augmented}

The Fisher matrix allows us to compute the expected errors on cosmological parameters for an ideal experiment. We now want to introduce the possibility of sources of systematic error. Suppose that there is some imperfectly-known aspect of the experimental setup that can be characterized by a parameter $p$. The experimenter has measured $p$ with some uncertainty, so that the probability density is sharply peaked around some measured value $\hat p$.
Often, we are willing to assume that the probability density $f(p|\hat p)$ is
Gaussian with mean $\hat p$ and known variance $\sigma_p^2$.

%
%

Let us include the experimental parameter $p$ along with the cosmological parameters $\vec\theta$ in the likelihood function:
\begin{align}
L(\vec\theta,p)&\equiv f(\vec r,p | \vec\theta,\hat p)
=f(\vec r | \vec\theta,p)f(p|\hat p)
\notag\\
&=\left({\exp\left(-{1\over 2}\vec r^T\Gammabold^{-1}\vec r\right)
\over (2\pi)^{n/2}{\rm det}^{1/2}(\Gammabold)}\right)\left(
{\exp(-(p-\hat p)^2/2\sigma_p^2)\over \sqrt{2\pi}\,\sigma_p}\right).
\end{align}
In this expression, the covariance matrix $\Gammabold$ depends on both the
cosmological parameters $\vec\theta$ and the experimental parameter $p$.


We want to determine the effect that our ignorance of $p$ has on our attempt to measure $\vec\theta$. We can do this by treating $p$ along with $\vec\theta$ as unknowns to be estimated simultaneously and calculating
an ``augmented'' Fisher matrix whose rows and columns correspond to the parameters $\vec\theta^{(a)}=(\theta_1,\theta_2,\ldots,\theta_m,p)$.

The log-likelihood is given by
\begin{align}
{\cal L}&\equiv -\ln L\notag\\
&={1\over 2}\left(\vec r^T\Gammabold^{-1}\vec r+\ln\det\Gammabold+{(p-\hat p)^2\over\sigma_p^2}
\right)+\mbox{const.}
\label{eq:loglaug}
\end{align}

In calculating the Fisher matrix elements $F_{jk}=-\langle\partial^2{\cal L}/
\partial\theta^{(a)}_j\partial\theta^{(a)}_k\rangle$, the ensemble average in principle includes an average over all possible values of $\hat p$ as well as over all possible values of $\vec r$.
However, it happens that all second derivatives of $\cal L$ with respect to ${\vec \theta}$ and $p$ are independent of $\hat{p}$,
so the averaging over $\hat p$ has no effect.
Thus, the upper left $m\times m$ block of the augmented Fisher matrix is still
given by equation (\ref{eq:finalfisher}). All that remains is to determine the final row and column.
%
%

For $j\le m$ we have
\begin{equation}
F_{j,m+1}={1\over 2}{\rm Tr}\left((\Gammabold^{(r)})^{-1}{\partial\Gammabold^{(r)}\over\partial
\theta_j}(\Gammabold^{(r)})^{-1}{\partial\Gammabold^{(r)}\over\partial p}\right).
\end{equation}
Note that this is just the usual Fisher matrix element, treating $p$ as a normal cosmological parameter. 
The very last term, in the lower right corner has an extra $1/{\sigma_p^2}$ in it
\begin{equation}
F_{m+1,m+1}={1\over 2}{\rm Tr}\left((\Gammabold^{(r)})^{-1}{\partial\Gammabold^{(r)}\over\partial p}
(\Gammabold^{(r)})^{-1}{\partial\Gammabold^{(r)}\over\partial p}\right)+
{1\over\sigma_p^2},
\end{equation}
arising from the term $(p-\hat p)^2/\sigma_p^2$ in equation (\ref{eq:loglaug}).

In the formulae above we have assumed that we only have one systematic parameter. The generalization to $n$ parameters is straightforward, leading to an $(m+n)\times (m+n)$ augmented Fisher matrix.

\subsection{Rotation errors}
\label{sec:rot1}

In section \ref{sec:results}, we consider two examples of systematic
errors: gain errors and rotation errors. In order to assess the effect of these errors on power spectrum estimates, we need to know their effects on the visibility covariance matrix. The case of gain errors is trivial: a gain error $g$ on a given antenna multiplies all of the micro-visibilities involving that antenna by $1+g$. The case of rotation errors requires more attention.

Suppose that each antenna $j$ has its polarizers rotated by some unknown angle $p_j$, with corresponding rotation matrix ${\bf R}_j={\bf R}(p_j)$. 
Consider a measurement of a micro-visibility pair $(V_{Q,(ab)},V_{U,(ab)})$,
where $(ab)$ labels an atenna pair. Then the
observed Stokes matrix ${\bf S}$ of (\ref{eq:stokes}) is related
to the true (error-free) matrix $\hat {\bf S}$ as
\begin{equation}
{\bf S}={\bf R}_a\hat {\bf S}{\bf R}_b^\dag.
\end{equation}
Because the visibilities are linearly related to the matrix ${\bf S}$, they obey a similar rule:
\begin{equation}
{\bf V}={\bf R}_a\hat {\bf V}{\bf R}_b^\dag
\end{equation}
Applying the rotation matrices to the Stokes visibility matrix,
we find
that
\begin{align}
V_{Q(ab)}&=\hat V_{Q(ab)}\cos(p_a+p_b)+\hat V_{U(ab)}\sin(p_a+p_b),\\
V_{U(ab)}&=-\hat V_{Q(ab)}\sin(p_a+p_b)+\hat V_{U(ab)}\cos(p_a+p_b).
\end{align}
That is, the visibilities $\begin{pmatrix}V_{Q(ab)}\\V_{U(ab)}\end{pmatrix}$ are simply multiplied by the rotation matrix corresponding to $(p_a+p_b)$.

Now consider the covariance between two pairs of \textit{macro}-visibilities $V_{XJ}$ and $V_{XK}$, where $X\in\{Q,U\}$ and $J,K$ label the visibilities.
We can express their covariances as a $2\times 2$ submatrix of the full covariance matrix:
\begin{equation} \label{macro cov}
{\bf C}=
\left< {\bf v}_J{\bf v}_K^\dag\right>,
\end{equation}
where
\begin{equation}
{\bf v}_J=\begin{pmatrix}V_{QJ}\\V_{UJ}\end{pmatrix}.
\end{equation}

In the absence of systematic errors, the covariance matrix 
is
\begin{equation} \label{c0}
{\bf C_0} ={1\over N_JN_K}\sum_{(ab)\in S_J}\sum_{(cd)\in S_K}
\left<{\bf v}_{(ab)}{\bf v}_{(cd)}^\dag\right>
\end{equation}

As we saw above, each micro-visibility is modified as follows:
\begin{equation}
{\bf v}_{(ab)}\to {\bf R}_a{\bf R}_b{\bf v}_{(ab)}.
\end{equation}

The $2\times 2$ covariance matrix corresponding to macro-visibilities $J,K$
is therefore
\begin{align}
{\bf C}&={1\over N_JN_K}\sum_{(ab)\in S_J}\sum_{(cd)\in S_K}
\left<{\bf v}_{(ab)}{\bf v}_{(cd)}^\dag\right>\notag\\
&={1\over N_JN_K}\sum_{(ab)\in S_J}\sum_{(cd)\in S_K}
{\bf R}_a{\bf R}_b{\bf C}_0{\bf R}_d^\dag {\bf R}_c^\dag\notag\\
&=
\left({1\over N_J}\sum_{(ab)\in S_J} {\bf R}_a{\bf R}_b\right)
{\bf C}_0
\left({1\over N_K}\sum_{(cd)\in S_K} {\bf R}_c{\bf R}_d\right)^\dag\notag\\
&=
{\cal R}_J{\bf C}_0{\cal R}_K^\dag,
\end{align}
where the matrices ${\cal R}$ are defined by the last line.

We can apply this rule to each pair of macro-visibilities to
determine the effect of the rotation errors on the entire
visibility covariance matrix.

\section{Approximate analytic treatment}
\label{sec:analytic}

Bunn \cite{bunnsys} gave an approximate analytic treatment of various sources
of systematic errors in CMB interferometry. We will compare
the full Fisher matrix calculation described above to this treatment.
In this section, we briefly review and extend the methods outlined in
that paper.

\subsection{Statistical errors on power spectrum estimates}

Before introducing systematic errors, we present a simple method
for estimating purely statistical errors on power spectrum
estimates.

Following \cite{bunnsys}, we begin by imagining an experiment that measures only a single visiblity pair $(V_Q,V_U)$.
According to equation (4.8b) of \cite{bunnsys}, for  a single
visibility pair, we can find an unbiased estimator of $C_l^B$ in a band centered on $l=2\pi u$,
\begin{equation}
\hat{C}_{B,\rm single}=(\pi\beam^2)^{-1}\gamma (\overline{c^2}\lvert V_U\rvert^2-\overline{s^2}\lvert V_Q\rvert^2),
\end{equation}
where $\beam$ is the beam size,
\begin{equation}
\gamma=[(\overline{c^2})^2-(\overline{s^2})^2]^{-1},
\end{equation}
and
\begin{align}
\overline{c^2}&=\frac{\int{\lvert\tilde{A^2}(\mathbf{k}+2\pi \mathbf{u}})\rvert^2\cos^2(\phi)d^2\mathbf{k}}{\int{\lvert\tilde{A^2}(\mathbf{k}+2\pi \mathbf{u}})\lvert^2 d^2\mathbf{k}}\\
\overline{s^2}&=\frac{\int{\lvert\tilde{A^2}(\mathbf{k}+2\pi \mathbf{u}})\rvert^2\sin^2(\phi)d^2\mathbf{k}}{\int{\lvert\tilde{A^2}(\mathbf{k}+2\pi \mathbf{u}})\lvert^2 d^2\mathbf{k}}=1-\overline{c^2}
\end{align}

We have defined $\theta_B$ as the ratio of measured to true power, so the estimator of this quantity is $\hat\theta_{B,\rm single}=\hat C_B/C_B$.
After some algebra, we find that the variance of $\theta_{B,\rm single}$ is
\begin{align}
\sigma_{B,\rm single}^2&=\langle\hat\theta_{B,\rm single}^2\rangle-\langle\hat\theta_{B,\rm single}\rangle^2
\notag\\
&=\gamma^2\left[2(\overline{c^2})^2(\overline{s^2})^2\left(\frac{C_E}{C_B}\right)^2+((\overline{c^2})^4+(\overline{s^2})^4)\right.\notag\\
           &\left.+(2(\overline{c^2})^3\overline{s^2}+2(\overline{s^2})^3\overline{c^2})\left(\frac{C_E}{C_B}\right)\right]
\label{eq:sbsingle}
\end{align}

We now consider an experiment with many measured macro-visibilities,
and
regard each macro-visibility pair as an independent measurement of $\theta_B$. Then the minimum-variance estimator from the entire data set is the usual weighted average, with weights given by the inverse variance, and the variance of this estimator is
\begin{equation}\label{oldberror}
\sigma_{B,\rm final}^2=\left(\sum_J \sigma_{BJ}^{-2}\right)^{-1},
\end{equation}
where $\sigma_{BJ}^2$ is given by (\ref{eq:sbsingle}) for baseline $J$.

In Section \ref{sec:results}, we will compare the results of this
simple approximation with the full Fisher matrix calculation,
as well as with the approximation in which calculate the fisher matrix while keeping only diagonal
elements of the (real) covariance matrix.

\subsection{Systematic errors}

We now summarize the key results of \cite{bunnsys}.
Once again, we begin by imagining measurements of the intensity
and polarization visibilities, ${\bf v}\equiv 
(V_I,V_Q,V_U)^T$ for a single
baseline. In many cases, the leading-order effect of
a systematic error is to ``mix'' the visibilities, 
inducing errors $\delta{\bf v}={\bf E}\cdot{\bf v}$ for
some mixing matrix ${\bf E}$.

\begin{figure}[t]
\includegraphics[width=3.5in]{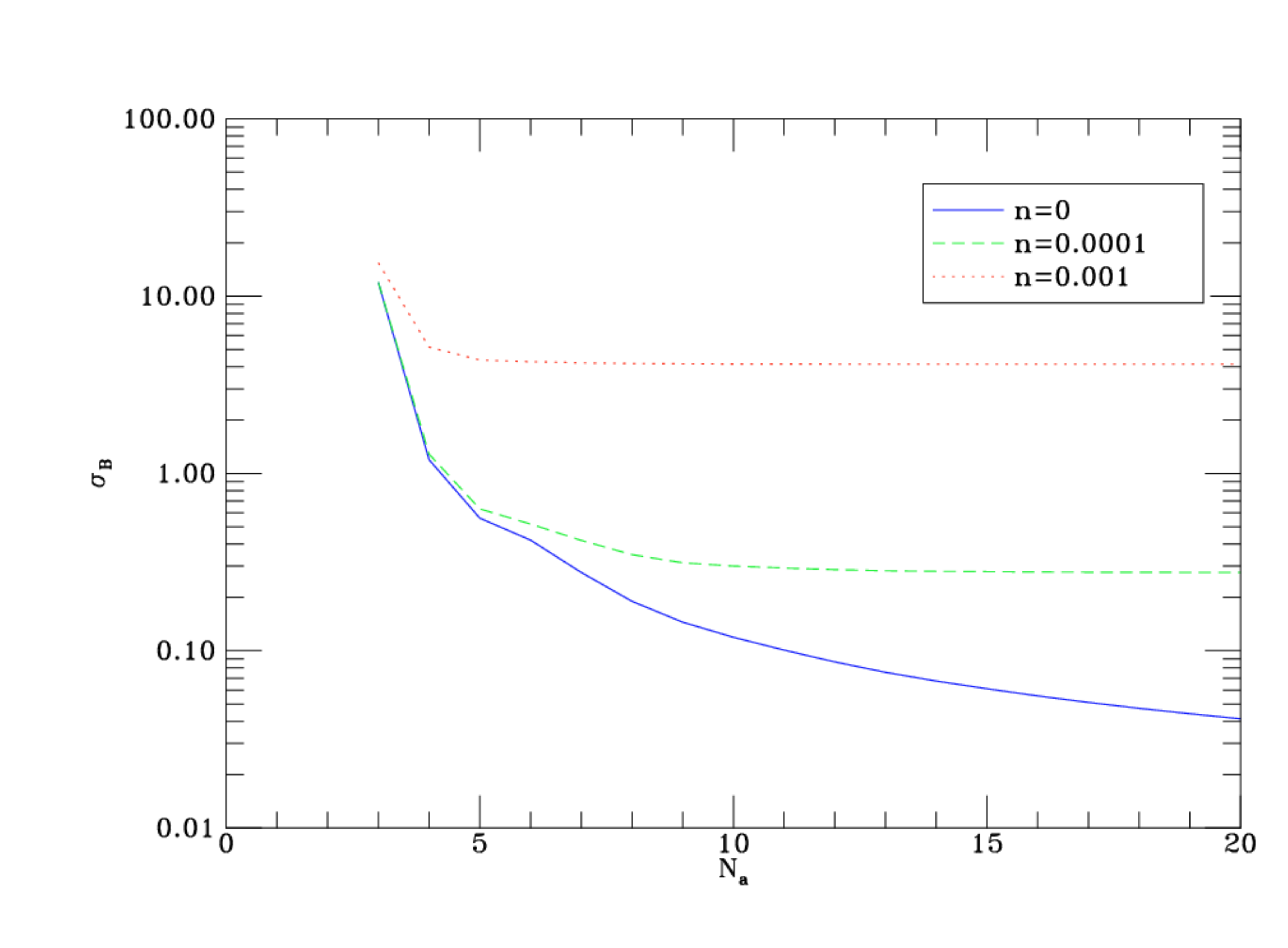}
\caption{The relation between $\sigma_B$ and $N_a$ for noise levels 0, 0.0001\,$\mu$K, and 0.001\,$\mu$K. The beam size $\beam=0.0873$ rad.}
\label{fig:noise}
\end{figure}

If the experimenter then uses the data set ${\bf v}$ to optimally
estimate the E and B power spectra, without accounting
for the effect of the systematic error, then the resulting estimates will
have errors given by
\begin{equation}
\left({\delta\hat{C}_{\rm rms}^K}\right)^2
=p^2\sum_{I,J}\kappa^2_{K,IJ}{C^IC^J},
\end{equation}
where $I,J,K\in\{T,X,E,B\}$ label the power spectra (temperature, 
TE correlation, E polarization, B polarization), $p$ is the
rms amplitude of the error under consideration,
and the coefficients $\kappa_{K,IJ}$ can be calculated from 
the error mixing matrix ${\bf E}$. Because there is a hierarchy
in power spectrum amplitudes, with $T>X>E>B$, one can generally
keep only the dominant term in this sum.

\begin{figure*}[t]
\includegraphics[width=3in]{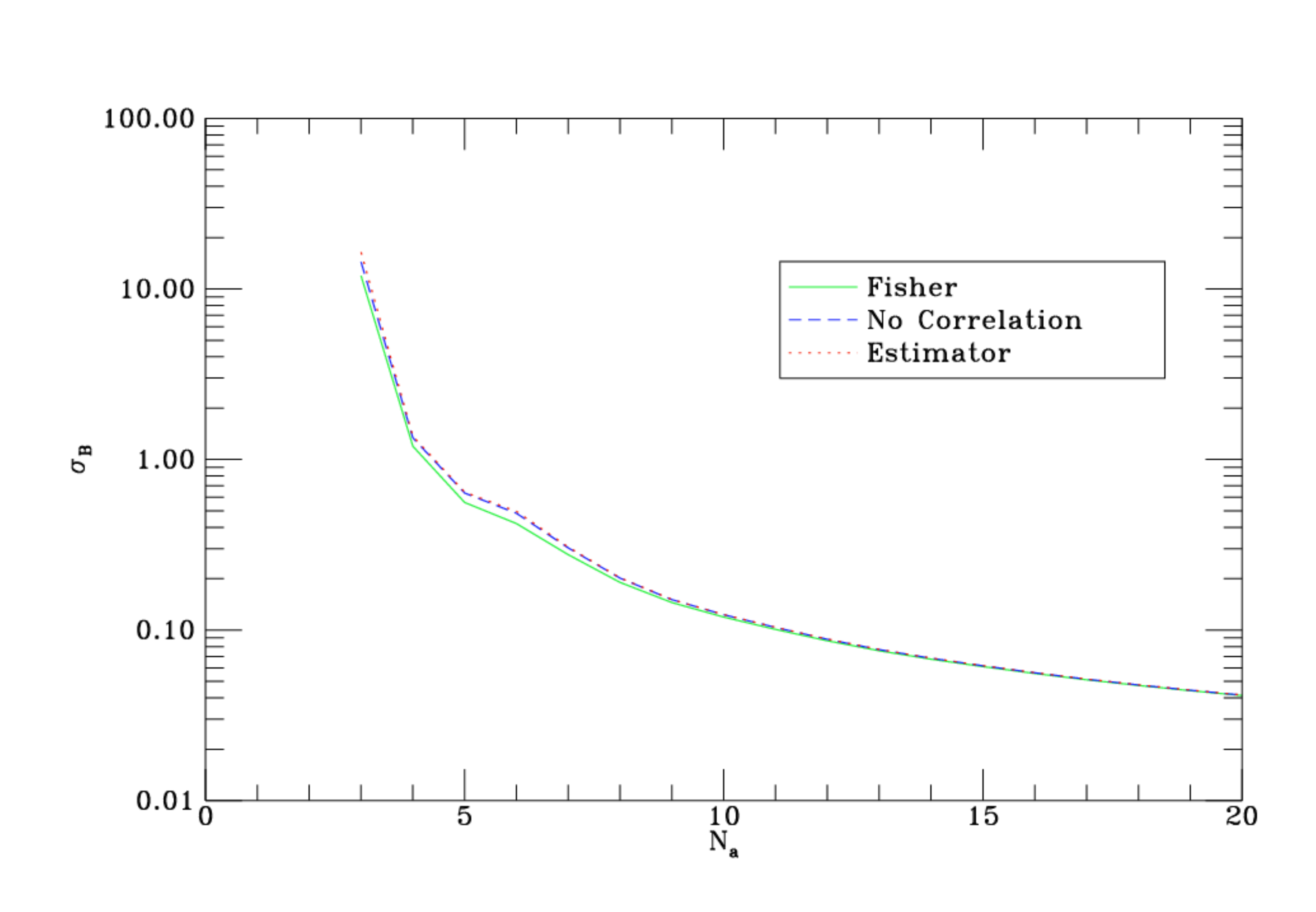}
\includegraphics[width=3in]{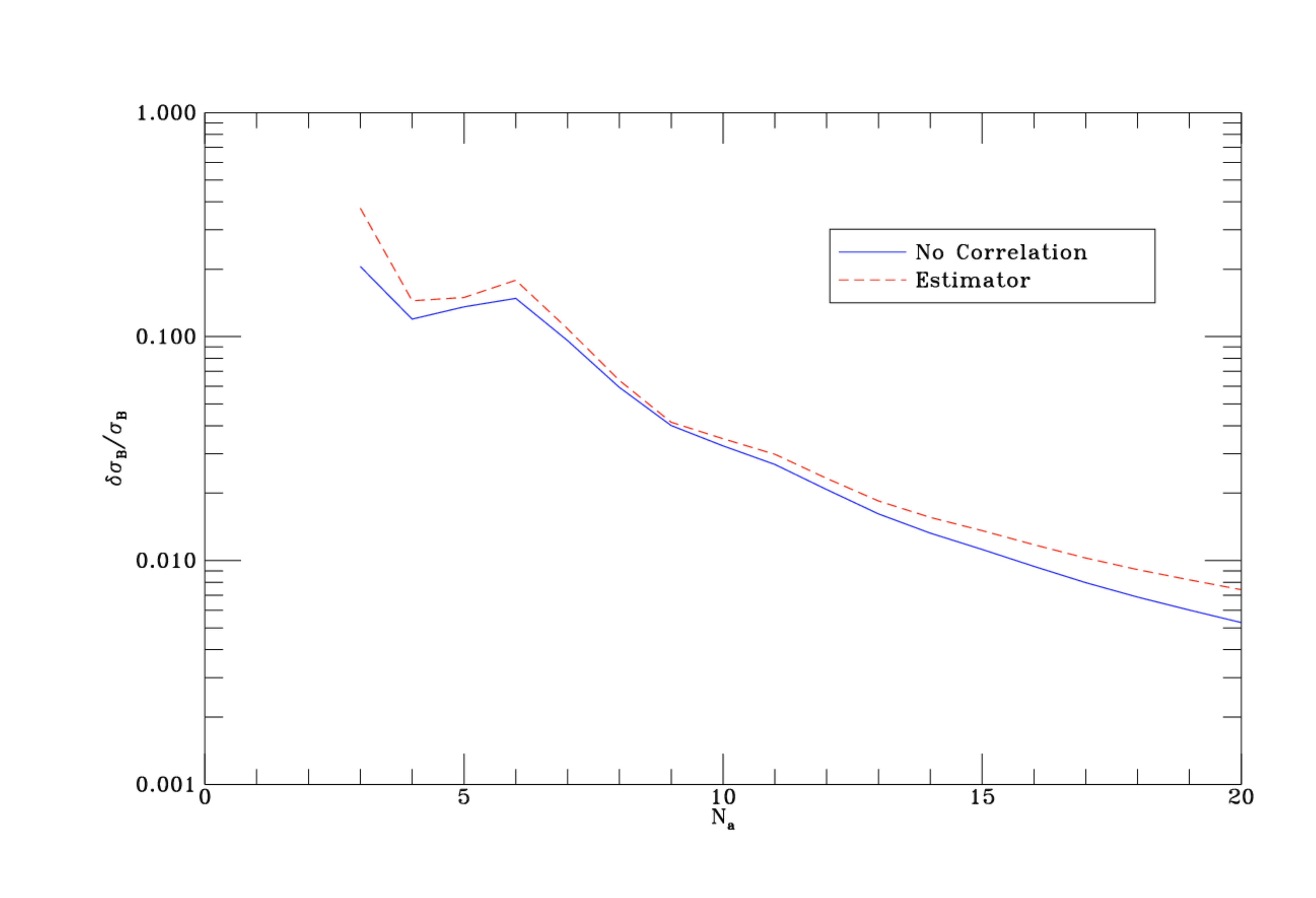}
\caption{The 
relationship between $\sigma_B$ and $N_a$
in the absence of systematic errors, calculated via the full Fisher matrix,
the Fisher matrix with correlations between distinct baselines neglected, and the analytic
estimator $\hat\theta_B$ of Section \ref{sec:analytic}. The left panel shows the errors calculated by the three
methods, and the right shows the difference (\ref{eq:deltasigmadef}) between
the approximations and the full Fisher calculations. Noise is set to zero in these calculations.}
\label{3comparison}
\end{figure*}

For example, for gain errors, the dominant term in the B power
spectrum error is the one that corresponds to contamination
by E power, with 
\begin{equation}
\kappa^2_{B,EE}=2\gamma^2(\overline{s^2}\,\overline{c^2})^2
\end{equation}

As in the case of statistical errors above, we can crudely estimate
the expected effect of a systematic error in a many-baseline experiment
by assuming that each baseline provides an independent power spectrum
estimate and performing an appropriate average.
See Section V of \cite{bunnsys} for details.

Equation (\ref{oldberror}) gives the analytic approximation for  $\sigma_B$ in the absence of systematic errors. In generalizing it to the case where a systematic error is included, we assume that, for each antenna, the error caused by the systematic error $p$ is independent of the statistical error,
so that  
\begin{equation}
\sigma_{Bj}^2=\sigma_{Bj,0}^2+\sigma_{Bj,p}^2.
\end{equation}
Here $\sigma_{Bj}$ is the error on the B power spectrum estimate corresponding to the $j$th micro-visibility. The quantity $\sigma_{Bj,0}^2$ is the noise variance for that visibility in the absence of systematic errors, given by equation (\ref{eq:sbsingle}), and $\sigma_{Bj,p}^2=(\delta\hat C^B_{\rm rms})^2/C_l^B$ is the additional variance induced by the systematic error. (Here $l=2\pi u$ for the given visibility.)
Because each macro-visibility is estimated via an average of the corresponding micro-visibilities, we can compute the expected error in our B power spectrum estimate for each macro-visibility. The final error on the B power is then given by equation (\ref{oldberror}).

\section{Results}
\label{sec:results}
We have calculated Fisher matrix errors for $N_a\times N_a$ square arrays
of close-packed antennas with Gaussian beam width $\beam = 5^\circ=0.0873$ 
rad, as in \cite{zhangml}.
We assume that the input power spectra are given by
the standard LCDM model \cite{planckparameters} with tensor-to-scalar ratio $r=0.1$, and that the experimenter's goal is to estimate the amplitudes of the E and B power spectra.
To be specific, let $C^E, C^B$ be the true power spectra. Then the parameters to be estimated are $\theta_E,\theta_B$ such that the estimated power spectra are
\begin{align}
\hat C^E&=\theta_EC^E\\
\hat C^B&=\theta_B C^B
\end{align}
Since $\theta_E,\theta_B$ are defined as relative amplitudes, their true value is one. The errors computed for these quantities will therefore be relative errors on the amplitude of the E and B signals. It would be straightforward to generalize these results to the estimation of multiple band powers in E and B.

We are particularly interested in the detectability of the faint B signal.
As a result, we will focus on the quantity
$\sigma_B=\sqrt{(F^{-1})_{BB}}$, which is the standard deviation of $\theta_B$ if none of the other parameters is known. 


\subsection{Exact Fisher Matrix with different noise levels}
We begin by ignoring all systematic errors and focusing on the effect of noise levels on the Fisher matrix. 
Assuming that all visibilities are subjected to white noise with
standard deviation $n$, the visibility covariance matrix is
\begin{equation}
\Gammabold_{noise}= \Gammabold+n^2\mathbf{I},
\end{equation}
where $\mathbf{I}$ is the identity matrix.

Figure \ref{fig:noise} shows the relationship between $\sigma_B$ and $N_a$ for  noise levels $n=0$, 0.0001, and 0.001 $\mu$K.\footnote{If these
values seem surprisingly small, note that, with our Fourier transform
and antenna pattern normalization conventions, the variance
of a visibility is $\sim \pi\beam^2C_{l}^E$.}
For noise-free experiments, $\sigma_B$ gradually decreases as $N_a$ increases. However, when some noise is present, there is a limit for $N_a$ at which $
\sigma_B$ cannot be attenuated even if we continue to increase the number of antennas.

For further analysis, we set $n=0.0002\,\mu$K, a value which allows for
a detection of B modes but for which noise contributions are still
significant.

\subsection{Comparison of calculation methods}

In this section, we compare errors calculated using the full Fisher matrix formalism with those based on two approximate methods. One is
the analytic approximations described in \ref{sec:analytic}. The
other is a Fisher-matrix calculation, with
the approximation that off-diagonal correlations between visibilities
can be neglected -- that is, that $\Gamma^{(r)}_{jk}=0$ for all $(j,k)$
corresponding to macro-visibilities with distinct baselines ${\bf u}_j\ne
{\bf u}_k$.

We begin by considering purely statistical errors and then introduce
systematic errors.

\subsubsection{Statistical errors}

We do not consider systematic errors in this section, meaning that our Fisher matrix will only be a $2\times 2$ matrix, corresponding to the two cosmological parameters $\theta_E,\theta_B$ to be estimated. One example of a typical Fisher matrix is 
\begin{equation}
F=\begin{pmatrix}
179 & 4.23\\
4.23& 3.08
\end{pmatrix}
\label{eq:fishexample}
\end{equation}
with inverse
\begin{equation}\label{eq:f-1}
F^{-1}=\begin{pmatrix}
0.00577 & -0.00792\\
-0.00792& 0.33586797
\end{pmatrix}
\end{equation}
with the parameters $N_a=10$, beam size $\beam=0.0873$ rad, and $n=0.0002\,\mu$K.
(We give extra significant figures for the $\sigma_B$ term for comparison
with later results.)
This matrix indicates, for instance, that with the given parameters
the least possible variance for measuring $\theta_B$ is $0.336$, relative to its mean value $1$.

Figure \ref{3comparison} compares the values of $\sigma_B$ calculated via the three methods. In these calculations, $\beam=0.0873$ as usual, but the noise level is set to zero. The left panel shows $\sigma_B\equiv\sqrt{(F^{-1})_{22}}$
for all three methods. For each of the two approximate methods, 
the right panel shows the fractional difference between the approximate
and full Fisher calculations:
\begin{equation}
{\delta\sigma_B\over\sigma_B}={\sigma_B^{\rm (approx)}-\sigma_B^{\rm (Fisher)}
\over\sigma_B^{\rm (Fisher)}}.
\label{eq:deltasigmadef}
\end{equation}


\begin{figure}[t]
\includegraphics[width=3in]{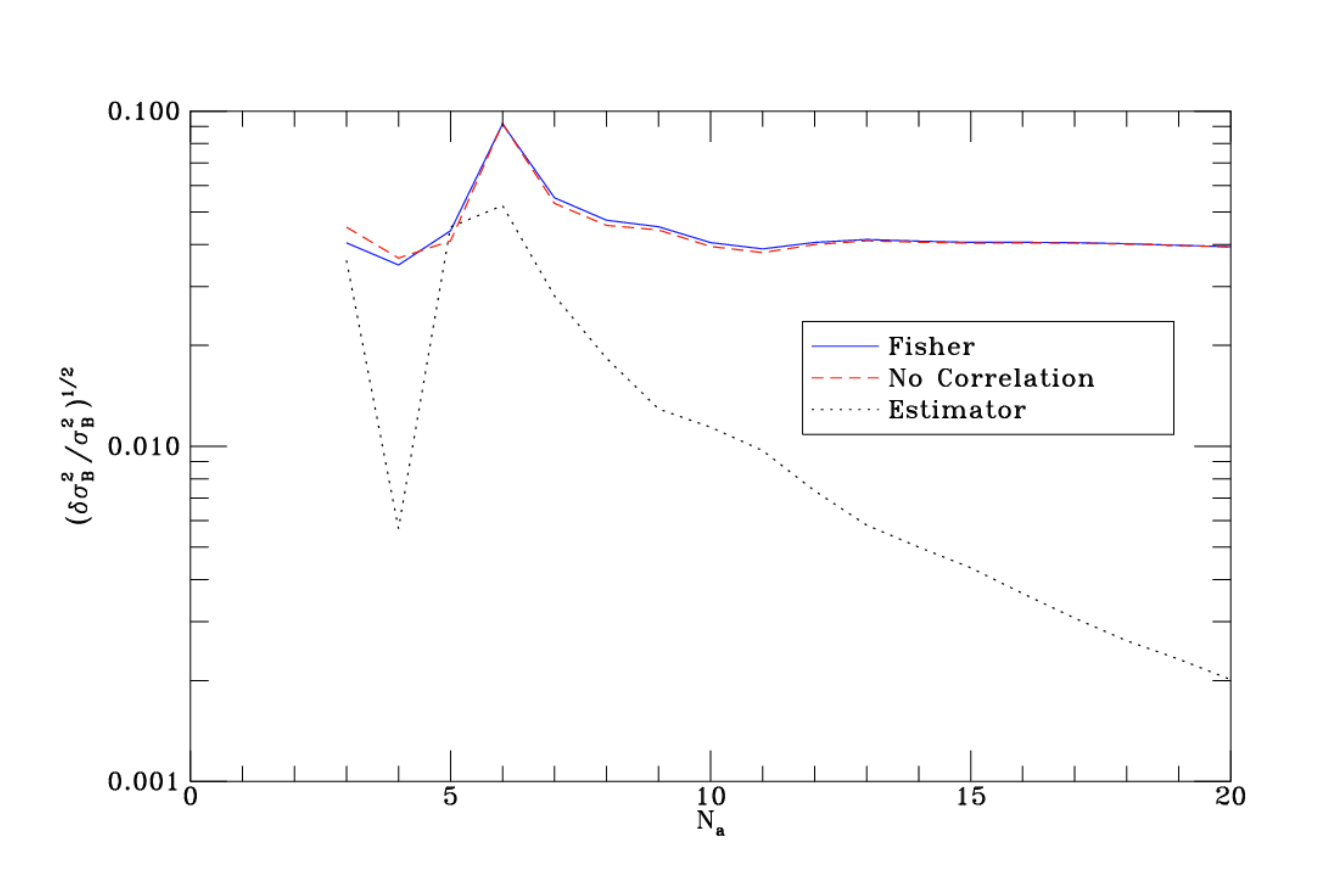}
\caption{The effect of a 10\% gain error on antenna 6, calculated via the three
methods listed in the previous figure. The quantity $\delta\sigma_B^2$ is the difference in variances of the error estimates with and without the gain error. Noise is set to zero in these calculations.}
\label{fig:3methodsgain}
\end{figure}

\begin{figure*}[t]
\includegraphics[width=3in]{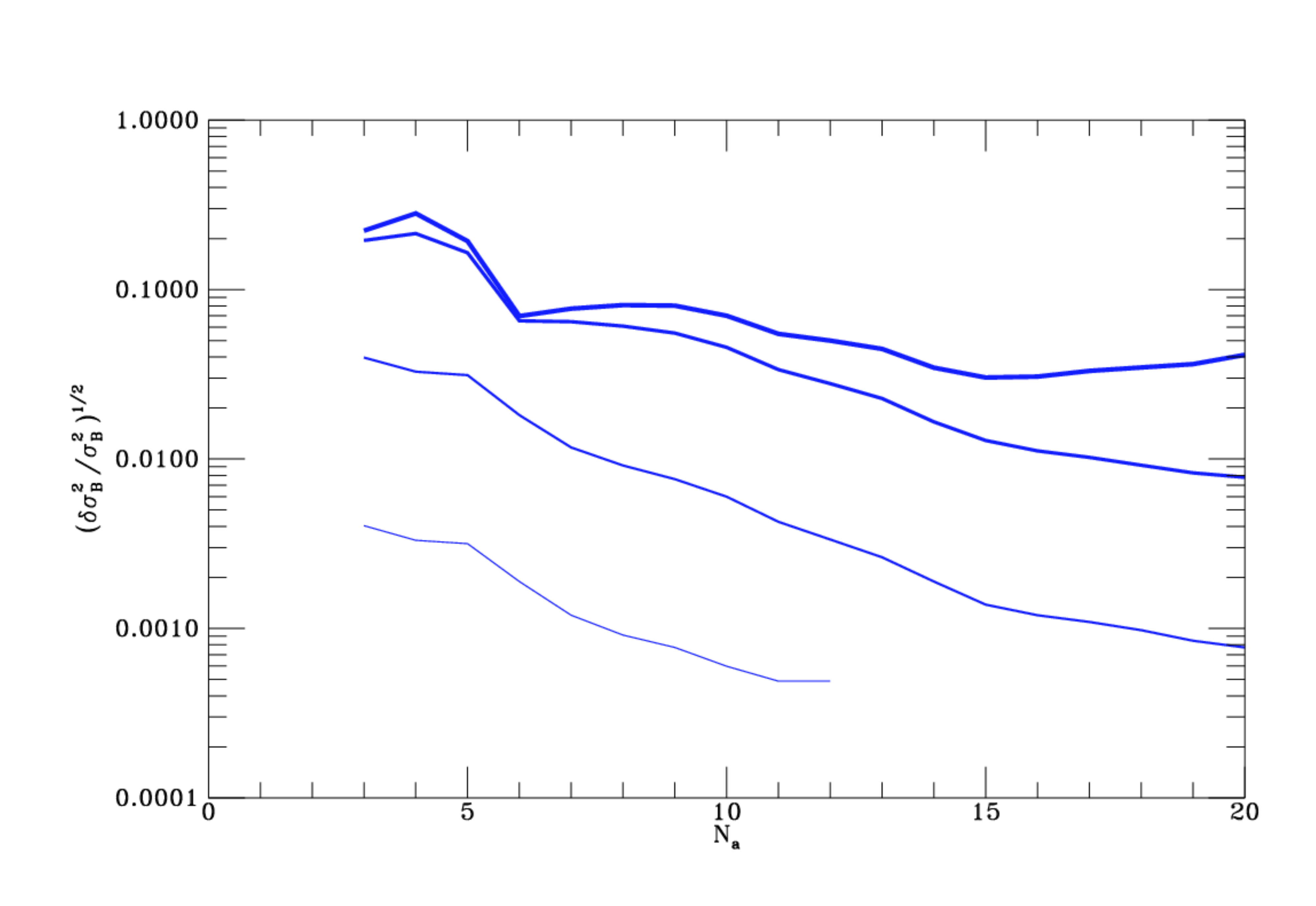}
\includegraphics[width=3in]{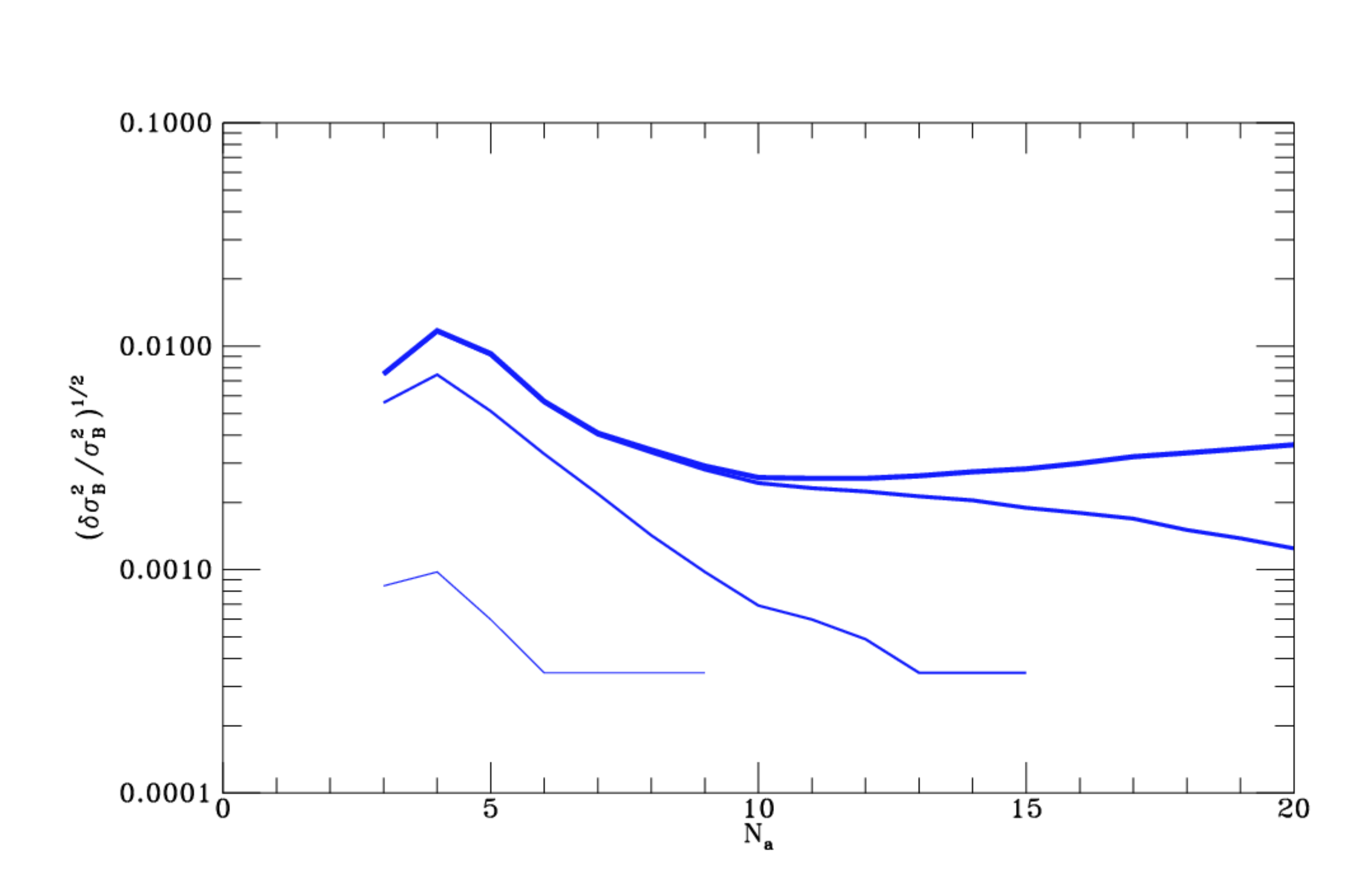}
\caption{
The additional error caused by a gain error in the sixth antenna (left) and a rotation error in the third antenna (right). From bottom to top, the curves correspond to error levels $\sigma_p=0.01,0.1,1,10$.
}
\label{fig:syserrresults}
\end{figure*}

\subsubsection{Systematic errors}

Suppose that there is a systematic error, such as a gain error or a rotation error, in one of the antennas of our interferometer. As described in Section \ref{sec:augmented}, we assume that this
error can be characterized by an unknown parameter $p$. For instance, in the case of a gain error, the unknown true gain is $(1+p)$ times the nominal gain, and for a rotation error $p$ is the angle by which the polarizers are misaligned. We treat $p$ as an additional parameter to be estimated, leading to a $3\times 3$ augmented Fisher matrix, whose rows and columns correspond to $\theta_E,\theta_B,p$. For example, if we introduce a 10\% gain error ($p=0.1$) to the sixth antenna of the array, the Fisher matrix (\ref{eq:fishexample})
and its inverse (\ref{eq:f-1}) are replaced by

\begin{align}
F_{\rm aug}&=\begin{pmatrix}
179   &    4.23   &    8.08\\
4.23  &    3.08   &    0.294\\
8.08  &    0.294  &    101
\end{pmatrix}\\
F_{\rm aug}^{-1}&=\begin{pmatrix}
0.00579 &   -0.00791 & -0.000439\\
-0.00791&    0.33588000 & -0.000346\\
-0.000439&  -0.000346   & 0.00993
\end{pmatrix}
\label{faug-1}
\end{align}
We see that the existence of this gain error increases the variance $\sigma_B^2$ for about $0.0036\%$.

Figure \ref{fig:3methodsgain} shows the effect of a 10\% gain error in the sixth antenna,
according to the three methods.
The quantity plotted is $(\delta\sigma_B^2/\sigma_B^2)^{1/2}$, where
$\delta\sigma_B^2=(\sigma_B^2)_{\rm sys}-(\sigma_B^2)_{\rm no\ sys}$ is the difference in the variances of the
estimates of B power with and without the systematic error.


\subsection{Gain and rotation errors}

Figure \ref{fig:syserrresults} shows sample results of the systematic error calculations based on the full Fisher matrix calculation.
The left panel shows the additional error induced by a gain error in the sixth antenna as a function of array size $N_a$, for four different values of the rms gain error $\sigma_p$. The right panel shows the same results for a rotation error in the third antenna. In this case, the error parameter $p$ is the angle through which the polarizer has been rotated in radians. In both cases, the noise level is set to 0.0002 $\mu$K.

In these sample results, we assumed an error that affected a single antenna, but
in a realistic experiment, we would expect all antennas to be affected.
The rigorous way to deal with this is to parameterize each of the systematic, build a large augmented Fisher matrix, and calculate $\sigma_B$ from
the resulting matrix. However, it is more convenient if we can treat each systematic error separately and add up the resulting $\sigma_B$'s.
The validity of this approach, of course, depends on the accuracy of treating
the individual errors as independent.

\begin{figure*}[t]
\includegraphics[width=3in]{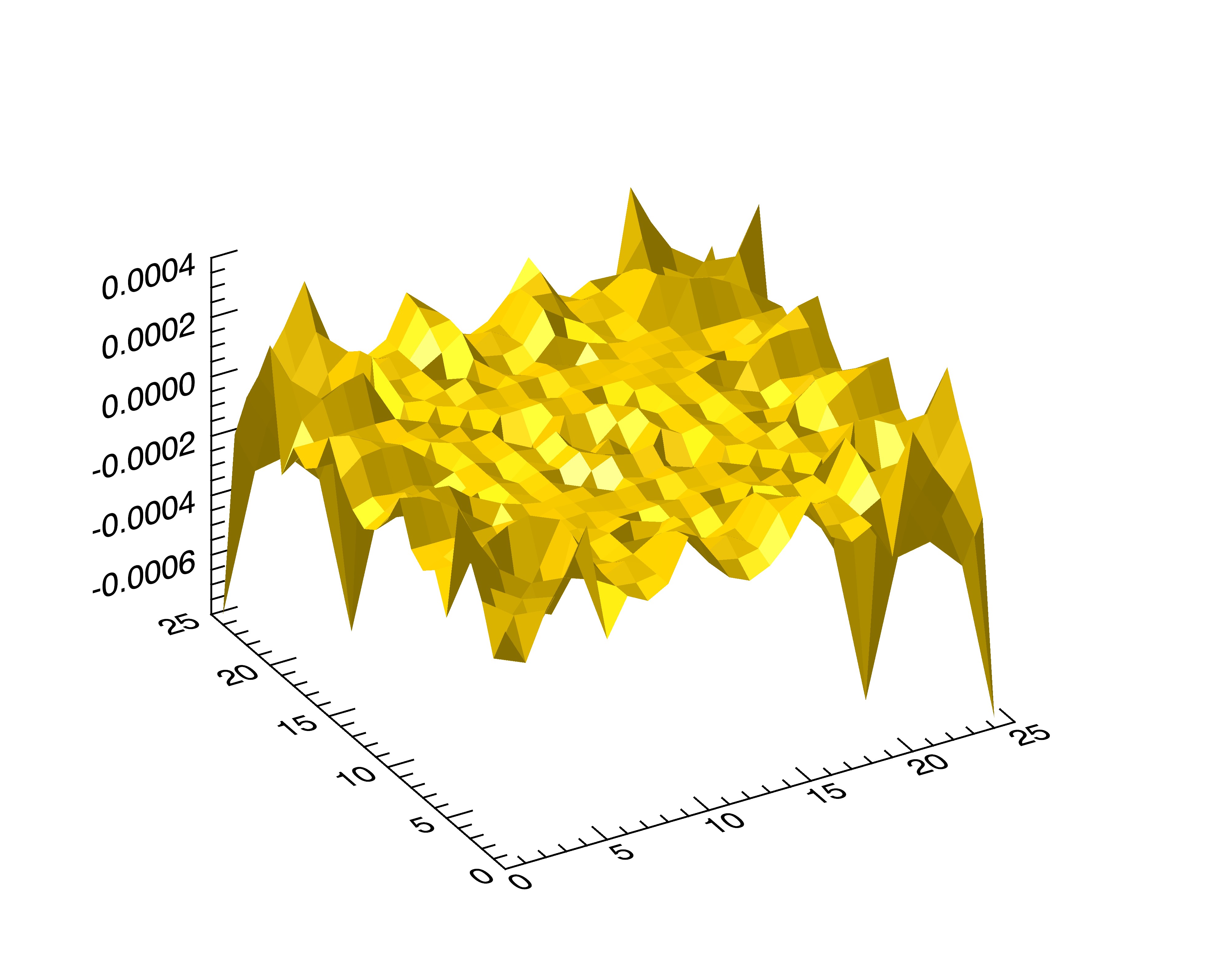}
\includegraphics[width=3in]{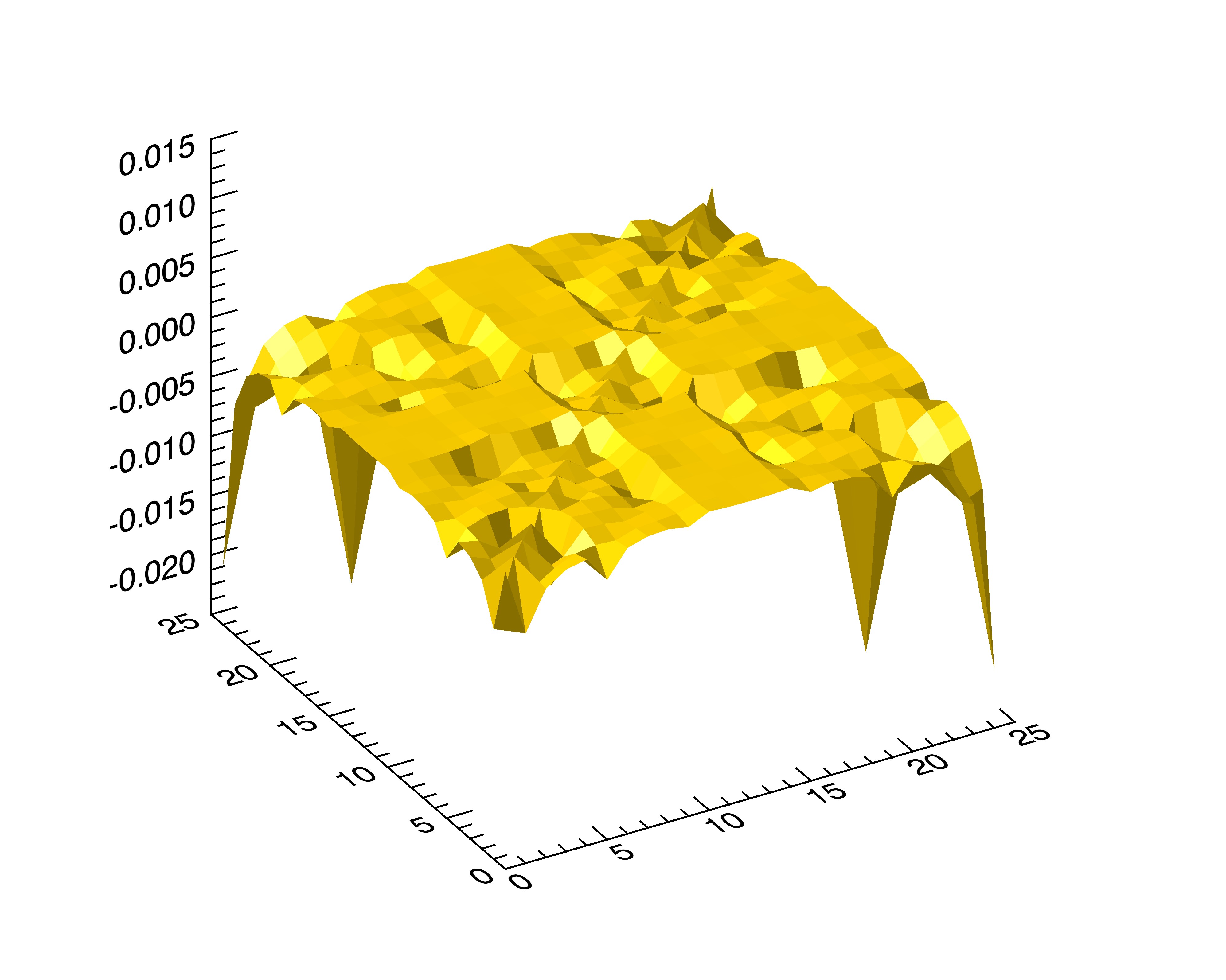}
\caption{
The inaccuracy caused by treating simultaneous errors in two antennas as if they were independent. The horizontal axes are the antennas of a $5\times 5$ array. The vertical axi is the quantity $\Delta_{ij}$ of equation (\ref{eq:combinederrors}).
}
\label{fig:combined}
\end{figure*}

Figure \ref{fig:combined} shows the results of tests of this independence assumption. For an array with $N_a=5$ and $n=0.0002\,\mu$K, we calculated the effects of a 10\% gain error in each of the 25 antennas. We also calculated the effect of simultaneous gain errors in each antenna pair. The quantity plotted is the difference between the simultaneous treatment and the quadrature sum of the individual errors, specifically
\begin{equation}
\Delta_{ij}\equiv{(\delta\sigma_B^2)_{ij}-((\delta\sigma_B^2)_i+
(\delta\sigma_B^2)_j)\over(\delta\sigma_B^2)_i+
(\delta\sigma_B^2)_j
},
\label{eq:combinederrors}
\end{equation}
where $i,j$ label antennas and $\delta\sigma_B^2$ is the additional 
variance induced by the given error. 

The right panel shows the equivalent results for a rotation error of 0.1 rad.

In both cases, the smallness of $\Delta_{ij}$ indicates
that treating the errors as independent is a good approximation, although the case of rotation errors is not as good as that of gain errors.

\section{discussion}
\label{sec:discussion}

We have presented a Fisher-matrix formalism for assessing the errors on power spectrum estimates from intereferometric measurements of CMB polarization, including the effects of systematic errors. This method occupies an intermediate position between crude analytic approximations \cite{bunnsys} and more accurate but far more computationally intensive end-to-end simulations \cite{karakcibayes,zhangml,karakcisys}. Although detailed simulations are necessary in the end, a fast semianalytic approach such as ours is likely to be useful during the early, exploratory phase of instrument design and optimization.

We have tested our method in a variety of ways, focusing particularly on systematic errors in antenna gain and in the orientation of polarizers of an antenna. Other sources of error can be treated in a similar way. In the absence of systematic errors, our calculations agree with those of simple analytic approximations. The agreement is less good when systematic errors are included (see Fig.~\ref{fig:3methodsgain}), which is not surprising given the crudeness of the approximations in the analytic method.

Most of our test results involve errors in a single antenna. Assuming different sources of error are independent, one expects the errors to add in quadrature. We verified for gain and rotation errors that this is a good assumption. In any case, it is not difficult to apply our formalism to the treatment of multiple errors simultaneously.

Because our tests were designed to illustrate the formalism, several idealization were made. We considered a single pointing of the instrument, without including sky rotation or mosaicking \cite{BW}. In addition, we assumed a single band power was to be estimated in each of the $E$ and $B$ power spectra. Finally, the errors we considered all involve only $E$-$B$ mixing, without contamination from the temperature anisotropy. There is no reason that our method cannot be extended to include these and other effects.

Although we developed the method for interferometric CMB polarimetry, the formalism developed herein can be applied to a variety of experiments. In particular, it would be very interesting to consider application to 21-cm tomography, which similarly involves extraction of small fluctuations signals from interferometric visibilities. Computationally expensive simulations are often done to assess the sensitivity of such instruments (\textit{e.g.}, \cite{zhang21}), but a faster semianalytic approach may be valuable as well.

\section*{Acknowledgments}
This work was supported by NSF Awards AST-1410133 and AST-0908319, and
by a Summer Fellowship from the University of Richmond School of Arts
and Sciences.

\bibliography{fim_paper}

\appendix

\section{The detailed form of the visibility covariance matrix}\label{appendix:covariance}

We assume a Gaussian antenna pattern as in equation (\ref{antennaPattern}),
\begin{equation*}
\hspace{2cm}A(\mathbf{r}) =  e^{-\mathbf{r}^2/{2\beam^2}}\hspace{3.5cm}(\ref{antennaPattern}),
\end{equation*} 
whose Fourier transform is
\begin{equation}
\tilde{A}(\mathbf{k}) = \tilde{A}^*(\mathbf{k}) = \frac{\beam^2}{2\pi}e^{-\beam^2\mathbf{k}^2/2},
\end{equation}
Substituting this into equation (\ref{vqvq}), we find
\begin{align}
& \qquad \langle V_Q(\mathbf{u}_1)V_Q^*(\mathbf{u}_2)\rangle \notag\\
&=(2\pi)^2\int{[P_E(\mathbf{k})\cos^2(2\phi)+P_B(\mathbf{k})\sin^2(2\phi)]}\notag\\
   &\qquad\tilde{A}^*(\mathbf{k}+2\pi \mathbf{u}_1) \tilde{A}(\mathbf{k}+2\pi \mathbf{u}_2) 
d^2 \mathbf{k}\hspace{2.3cm}(\ref{vqvq})\notag
\end{align}

Then
\begin{align}
&\qquad \langle
V_Q(\mathbf{u}_1)V_Q^*(\mathbf{u}_2)\rangle \notag\\
&=\beam^4\int{[P_E(\mathbf{k})\cos^2(2\phi)+P_B(\mathbf{k})\sin^2(2\phi)]}\notag\\
&\qquad \exp\left[-\frac{1}{2}\beam^2\left[(\mathbf{k}+2\pi \mathbf{u}_1)^2+(\mathbf{k}+2\pi \mathbf{u}_2)^2\right]\right] 
d^2 \mathbf{k}\notag\\
&=\beam^4\int{[P_E(k)\cos^2(2\phi)+P_B(k)\sin^2(2\phi)]}\notag\\
&\qquad \exp\big[-\beam^2\left[2\pi^2(u_1^2+u_2^2)+k^2\right]-2\pi\beam^2\notag\\
&\qquad\left[(u_{1x}+u_{2x})k\cos\phi+(u_{1y}+u_{2y})k\sin\phi\right]\big]kdkd\phi
.
\end{align}

For convenience, we make the following definitions:
\begin{align}\label{s1}
&S_1(k)=\int_0^{2\pi} \cos^2(2\phi)\notag\\
&\exp\left[-2\pi\beam^2\left[(u_{1x}+u_{2x})k\cos\phi+(u_{1y}+u_{2y})k\sin\phi\right]\right]d\phi,
\end{align}
\begin{align}\label{s2}
&S_2(k)=\int_0^{2\pi} \sin^2(2\phi)\notag\\
&\exp\left[-2\pi\beam^2\left[(u_{1x}+u_{2x})k\cos\phi+(u_{1y}+u_{2y})k\sin\phi\right]\right]d\phi,
\end{align}
\begin{align}\label{s3}
&S_3(k)=\int_0^{2\pi} \sin(4\phi)\notag\\
&\exp\left[-2\pi\beam^2\left[(u_{1x}+u_{2x})k\cos\phi+(u_{1y}+u_{2y})k\sin\phi\right]\right]d\phi,
\end{align}
and
\begin{equation}\label{qx}
q(x)=\exp\left[-\beam^2[2\pi^2(u_1^2+u_2^2)+k^2]\right].
\end{equation}
Then one can check that 
\begin{align}\label{vqvqf}
\langle V_Q(\mathbf{u}_1)V_Q^*(\mathbf{u}_2)\rangle &=
\beam^4\int_0^\infty P_E(k)q(k)S_1(k)kdk\notag\\
&\quad +\beam^4\int_0^\infty P_B(k)q(k)S_2(k)kdk.
\end{align}
Similarly, for the $U$-$U$ and $Q$-$U$ covariance matrix elements, 
\begin{align}\label{vuvuf}
\langle V_U(\mathbf{u}_1)V_U^*(\mathbf{u}_2)\rangle&=
\beam^4\int_0^\infty P_B(k)q(k)S_1(k)kdk\notag\\
&\quad+\beam^4\int_0^\infty P_E(k)q(k)S_2(k)kdk.
\end{align}
\begin{align}\label{vqvuf}
\langle V_Q(\mathbf{u}_1)V_U^*(\mathbf{u}_2)\rangle
&=-\frac{1}{2}\beam^4\int_0^\infty P_E(k)q(k)S_3(k)kdk\notag\\
&\quad+\frac{1}{2}\beam^4\int_0^\infty P_B(k)q(k)S_3(k)kdk.
\end{align}

Equations (\ref{vqvqf}), (\ref{vuvuf}), and (\ref{vqvuf}), combined with equations (\ref{s1}), (\ref{s2}), (\ref{s3}), and (\ref{qx}), are sufficient to present a workable algorithm to calculate the visibility covariance matrix. We now provide the analytic forms of the integrals $S_1$, $S_2$, and $S_3$ as the ending of this section. We will take $S_1$ as an example and directly give $S_2$ and $S_3$.

Let 
\begin{align}
a&=(u_{1x}+u_{2x})k,\\
b&=(u_{1y}+u_{2y})k.
\end{align}

Then 
\begin{equation}
S_1(k)=\int\cos^2(2\phi)\exp\left[-2\pi\beam^2(a\cos\phi+b\sin\phi)\right]d\phi.
\end{equation}
Now let
\begin{align}
t&=\arctan\frac{b}{a},\\
\theta&=\phi+t,\\
l&=\sqrt{a^2+b^2}.
\end{align}
Then
\begin{align}
S_1(k)&=\int\cos^2(2\theta-2t)\exp\left[-2\pi\beam^2l\sin\theta\right]d\theta\notag\\
&=\int\cos^2(2\theta)\cos^2(2t)\exp(-2\pi \beam^2l\sin\theta)d\theta\notag\\
&+\int\sin^2(2\theta)\sin^2(2t)\exp(-2\pi \beam^2l\sin\theta)d\theta.
\end{align}

Define 
\begin{align}
S_{11}(k)&=\int\cos^2(2\theta)\exp(-2\pi \beam^2l\sin\theta)d\theta,\\
S_{12}(k)&=\int\sin^2(2\theta)\exp(-2\pi \beam^2l\sin\theta)d\theta.
\end{align}
Then 
\begin{equation}
S_1(k)=S_{11}(k)\cos^2(2t)+S_{12}(k)\sin^2(2t).
\end{equation}

The integrals $S_{11}$ and $S_{12}$ can be written analytically using the modified Bessel function of the first kind $I_n(x)$:
\begin{align}
S_{11}(k)&=\frac{2}{l^3\pi^2\beam^6}\big[l\pi\beam^2(3+l^2\pi^2\beam^4)I_0(2l\pi\beam^2)\notag\\
&-(3+2l^2\pi^2\beam^4)I_1(2l\pi\beam^2)\big],
\end{align}
and
\begin{equation}
S_{12}(k)=\frac{2}{l^2\pi\beam^4}\big[2l\pi\beam^2I_1(2l\pi\beam^2)-3I_2(2l\pi\beam^2)\big].
\end{equation}

We can follow the similar reasoning to calculate $S_2$ and $S_3$. The results are
\begin{equation}
S_2(k)=S_{11}(k)\sin^2(2t)+S_{12}(k)\cos^2(2t),
\end{equation}
and
\begin{align}
S_3(k)&=-\frac{2\sin(4t)}{l^3\pi^2\beam^6}\big[l\pi\beam^2(6+l^2\pi^2\beam^4)I_0(2l\pi\beam^2)\notag\\
&-2(3+2l^2\pi^2\beam^4)I_1(2l\pi\beam^2)\big].
\end{align}



\end{document}